\documentstyle[preprint,aps]{revtex}
\begin{document}
\title{Microwave--Driven SET--Box with Superconducting Island}

\author{A. H\"adicke and W. Krech}
\address{Institut f\"ur Festk\"orperphysik,
        Friedrich-Schiller-Universit\"at Jena,
        Max-Wien-Platz 1\\
        D--07743 Jena,
        Germany}
\date{June 1994}
\maketitle
\begin{abstract}
The effect of a high--frequency microwave irradiation on the averaged
stationary charge in a single--electron tunneling box (SET--box)
with superconducting island is investigated.
The resulting $\langle Q\rangle$--$Q_g$ characteristics are modified
compared with those from the well--known constant voltage--driven box.
It can be shown that there is a cross--over from the 2e--periodic regime
in $Q_g$ corresponding to the parity effect to
1e--periodic behavior not only with
increasing temperature but also with increasing irradiation amplitude.
\end{abstract}
\pacs{73.40G, 73.40R, 05.60}
%
\section{Introduction}
\label{kap1}
The tunneling of single electrons (SET) through ultrasmall junctions
has become of much experimental and theoretical interest
(see e.g.~\cite{z1})).
The main conditions to have charging effects (Coulomb blockade) are
low temperatures ($k_BT\ll E_C$) and large tunnel resistances
($R_T\gg R_Q=h/e^2=25.8\:k\Omega$). This guarantees that single--electron
effects are lifted out of thermal and quantum fluctuations.
$e$ and $h$ are the elementary charge and Planck's constant, respectively,
$k_B$ is Boltzmann's constant. $E_C$ denotes the relevant Coulomb energy
under consideration.

We consider the so called SET--box formed by a SET junction and a
capacity $C_s$ in series as described in Fig.~1.
It is known that the mean electric charge on the island of this
SET--box can be increased in integer steps by a continuous
increase of the external voltage $V=V_o$ (see e.g.~\cite{est1})
corresponding to an increase of the charge $Q_g$ at the capacity $C_s$
($Q_g=C_sV_o$). The staircase function $\langle Q\rangle$ vs. $Q_g$
is the sharper the temperature is lower (see Fig.~2).
The tunnel junction is characterized by the capacity $C$ and the
imaginary part of the quasi--particle current amplitude Im$I_q$ which
is related with the phenomenological tunnel resistance $R_T$.

A new aspect which has thoroughly been investigated as well
theoretically and experimentally are parity effects on the superconducting
island~[3--11].
It has been observed
that the parity based 2e--periodicity of the charge characteristics
disappears in the vicinity of a cross--over temperature which is much lower
than the critical temperature of the superconductor.

There are some articles dealing with time--dependent phenomena in
mesoscopic tunnel junctions~[12--16].
We have investigated in an earlier paper~\cite{kre9} the influence of a
microwave irradiation on the stationary behavior of a normalconducting
double--junction.

The aim of this paper is to investigate the influence of a microwave
irradiation on SET--box parity effects in stationary regime.
Because of the fact that in some sense the microwave irradiation acts
similarly like temperature increase the cross--over mentioned above can
also be achieved by a sufficient strong irradiation.

\section{Normalconducting Box}
\label{kap2}

For a normalconducting island the relevant energy function which enters the
transition rates reads as
\begin{equation}
\label{energfkt0}
E^{ch}_n(Q_g)=\frac{e^2}{2(C+C_s)}(n-Q_g/e)^2\;.
\end{equation}
$n$ labels the number of electrons on the island. The index ``ch'' means
that this energy is the charging energy.
The charge $Q_g$ at the capacity $C_s$ is connected with the external voltage
$V_o$ via the relation
\[Q_g=V_oC_s.\]

The master equation describing the dynamics of the occupation probabilities
of the SET--box reads as~\cite{ave2,kre3}
\begin{equation}
\label{master0}
\frac{d}{dt}\sigma_n(t)=-(\Gamma_{n-1,n}+\Gamma_{n+1,n})\sigma_n(t)
+\Gamma_{n,n+1}\sigma_{n+1}(t)+\Gamma_{n,n-1}\sigma_{n-1}(t)\;.
\end{equation}
The rates $\Gamma_{n\pm 1,n}$ which depend on the energy differences
(cf.\ Eq.~(\ref{energfkt0}), $E_C=e^2/(2(C+C_s))$)
\begin{equation}
\label{energdiff}
\Delta E^{ch}_{n\pm 1,n}(Q_g)
=E^{ch}_{n\pm 1}(Q_g)-E^{ch}_n(Q_g)=E_C\left(1\pm 2n\mp 2Q_g/e\right)\;,
\end{equation}
are given by
\begin{equation}
\label{rates0}
\Gamma_{n\pm 1,n}(Q_g)=
\frac{1}{e}\mbox{Im}I_q[-\Delta E^{ch}_{n\pm 1,n}(Q_g)/\hbar]
\left\{1-\exp[\Delta E^{ch}_{n\pm 1,n}(Q_g)/k_BT]\right\}^{-1}\;.
\end{equation}
$I_q$ are the quasi particle current amplitudes of the NIN junction. The
rates $\Gamma_{n\pm 1,n}$ correspond to the tunneling processes
connected with the charge transitions $n\rightarrow n\pm 1$, respectively.
The succession of indices of the rates is the same as for the energy
differences, the first index refers to the final state and the second to the
initial state. The mean stationary charge in the box is given by
\begin{equation}
\label{charge0}
\langle Q\rangle=e\sum\limits_n n\bar{\sigma}_n(Q_g)\;,
\end{equation}
where $\bar{\sigma}_n$ are the stationary solution of Eq.~(\ref{master0}).
They can be written as~\cite{amm1,seu1}
\begin{equation}
\label{sol}
\bar{\sigma}_n(Q_g)=\frac{1}{\cal{N}}\prod\limits_{m=-\infty}^{n-1}
\Gamma_{m+1,m}(Q_g)\prod\limits_{m'=n+1}^{\infty}\Gamma_{m'-1,m'}(Q_g)\;.
\end{equation}
${\cal N}$ guarantees the correct normalization. It can be shown
that the mean charge~(\ref{charge0}) is given by a Boltzmann
distribution~\cite{est1}
\begin{equation}
\label{charge1}
\langle Q\rangle =e\frac{\sum\limits_n n e^{-E^{ch}_n(Q_g)/k_BT}}{\sum\limits_n
e^{-E^{ch}_n(Q_g)/k_BT}}\;.
\end{equation}
This Boltzmann distribution shows that the
charge in the box corresponds to a stationary state and dynamical properties
expressed by the current amplitudes do not play any role.

\section{Normalconduction Box with Irradiation}
\label{kap3}

The influence of a microwave irradiation on single--electron tunneling through
single--junctions has been studied in Refs.~[13--16].
The microwave irradiation is described by an additional oscillating
voltage part
\begin{equation}
\label{voltage}
V(t)=V_o+V_1\cos\bar{\omega}t\;.
\end{equation}
In this way the microwave field is treated classically
neglecting possible back reactions due to tunneling.
This is reasonable because all calculations are only done in first
order perturbation theory.
The effect of a microwave irradiation in case of the box is very similar to
that of the double--junction~\cite{kre9} because there is the island which can
only be charged in integer quantum units.

But now the time averaged charge does not more satisfy a Boltzmann
distribution.
It is rather given by formula~(\ref{charge0}) where the transition rates
have been modified (cf. Eq.~(\ref{rates0}))
$\Gamma\longrightarrow \bar{\Gamma}$.
The new rates read as
\begin{equation}
\label{rates1}
\bar{\Gamma}_{n\pm 1,n}(Q_g)=\frac{1}{e}\sum\limits_rJ_r^2(a)
\mbox{Im}I_q[-\Delta E^{ch}_{n\pm 1,n}(Q_g)/\hbar-r\bar{\omega}]
\left\{1-e^{(\Delta E^{ch}_{n\pm
1,n}(Q_g)+r\hbar\bar{\omega})/k_BT}\right\}^{-1}\;.
\end{equation}
The parameter $a$ is given by
\[a=\frac{eV_1}{\hbar\bar{\omega}}\]
and $J_i$ are the Bessel functions of the first kind.
The description via a Boltzmann distribution fails because the irradiation
does not allow a pure stationary description. Also the
stationary parts of the occupation probabilities carry a relic of the
dynamics. From the mathematical point of view
the reason is the sum structure of the rates which does not
allow a factorization.
This approach is based on the assumption that possible measurements of the
island charge (e.g.\ by a on--chip SET--electrometer) would only
detect the time--averaged charge. Therefore, these rates correspond already
to an averaging with respect to the oscillations. But nevertheless they
contain information from the irradiation. For vanishing AC amplitudes these
expressions lead to the ordinary box transition rates
without irradiation~(\ref{rates0}).

The result is (see Fig.~3) that the increase of the AC voltage amplitude
gives rise to a smoothing of the mean charge function compared with that
for $V_1=0$. In this way the oscillating driving voltage smears out the
hard staircase and acts similarly like a temperature increase. This can
be demonstrated in the following manner. For $T=0$ the rates without
irradiation are nonvanishing only for $\Delta E^{ch}_{\ldots}<0$.
But $T>0$ there are also transitions if $\Delta E^{ch}_{\ldots}>0$.
In case of irradiation and $T=0$ the condition for nonvanishing rates is
$\Delta E^{ch}_{\dots}+r\hbar\bar{\omega}<0$
which means that there are also transitions for $\Delta E^{ch}_{\ldots}>0$.
The comparison of Figs.~2 and 3 shows that there is nearly no difference
for a specific choice of parameters.
The mechanism can be understood as photon assisted tunneling via excited
states. The main effect is due to 1--photon processes ($r=\pm 1$) because
multiple photon processes are strongly damped.
But the specific shape of the charge function depends on a
complicated interplay of the parameters $a$, $\bar{\omega}$ and $T$.

\section{Superconducting Box}
\label{kap4}
The phenomenological treatment of the superconducting case starts with
the observation that the energy content of the island depends on the parity
$P_n$ of the electron number $n$ on it. This means that for odd $n$ there
is the additional energy gap $\Delta$ in the energy function otherwise there
is no gap for even $n$ , $P_n=n\;\mbox{mod}\;2$ .
This corresponds to the substitution
\begin{equation}
\label{energfkt1}
E^{ch}_n(Q_g)\;\longrightarrow \;E_n(Q_g)=E^{(ch)}_n(Q_g)+\Delta\cdot P_n\;.
\end{equation}
The same approach as in Sec.~(\ref{kap2}) can be done using the energy
function~(\ref{energfkt1}). This consideration is correct for zero
temperature. Then the charge in the box jumps to the next integer value
at the crossing point of two neighbouring $E_n(Q_g)$ parabolas (see Fig.~4).
The shape of the resulting charge staircase function depends on the relation
$\lambda$ between the energy gap $\Delta$ and the Coulomb energy $E_C$
($\lambda=\Delta/E_C$) (see Fig.~5).
The $\langle Q\rangle$--$Q_g$ characteristics shows for $\lambda>0$ a
2e--periodic behavior in $Q_g$.
But this approach cannot be correct for $T>0$ because it does not explain
the cross--over from the 2e--periodicity to 1e--periodicity at a cross--over
temperature $T^{\ast}$ known from experiments~\cite{tuo1}. This cross--over
arises due to a temperature dependent difference of the free energies
$\delta F(T)$ of the island in the even and odd state~\cite{tuo1,jan1}.
$\delta F$ is reduced from $\Delta$ at $T=0$ by entropy contributions.
It is not an effect of the temperature dependence of $\Delta$ itself
because this temperature is still much lower than the critical temperature
$T_C$. Therefore $\Delta=\Delta(0)$ throughout this paper. A first order
approximation for $\delta F(T)$ reads as
\begin{equation}
\label{free}
\delta F(T)\approx\Delta -k_BT\ln N_{eff}\;.
\end{equation}
$N_{eff}$ is the effective number of states avaiable for excitation. It
depends only weakly on temperature
\[N_{eff}=2 N_{(n)}(0){\cal V}\Delta e^{\Delta/k_BT} K_1(\Delta/k_BT)\;.\]
$N_{(n)}(0)$ is the normal density of states at the Fermi level and ${\cal V}$
corresponds to the island volume. $K_1$ labels the modified Bessel function.
The cross--over temperature is defined by
\[\delta F(T^{\ast})=0\;.\]
For their specific sample Tuominen et al.~\cite{tuo1} found
\[\delta F(T)\approx \Delta (1-\frac{T^{\ast}}{T})\] with $T^{\ast}\approx 300
mK$
much less than $T_C$. Of course this is reasonable only for
$0\le T\le T^{\ast}$, the function $\delta F(T)$ is positive definite.
For $0<Q_g/e<(1+\lambda)/2$ and $k_BT<E_C$ one can use the approximation that
the system is mainly governed by the two states $n=0$ and $n=1$ (cf.~Fig.~4).
Generalizing the approach of Sec.~(\ref{kap2})
the stationary occupation probabilities obey the Boltzmann distribution
\begin{eqnarray}
\bar{\sigma}(0)&=&\frac{e^{-E^{ch}_0/k_BT}}{e^{-E^{ch}_0/k_BT}+
e^{-(E^{ch}_1+\delta F(T))/k_BT}}\;,\nonumber\\
\bar{\sigma}(1)&=&\frac{e^{-(E^{ch}_1+\delta F(T))/k_BT}}{e^{-E^{ch}_0/k_BT}+
e^{-(E^{ch}_1+\delta F(T))/k_BT}}\;.
\end{eqnarray}
Then $\langle Q\rangle$ is given by
\begin{equation}
\label{charge2}
\langle Q\rangle =e\frac{e^{-(E^{ch}_1(Q_g)
+\delta F(T))/k_BT}}{e^{-E^{ch}_0(Q_g)/k_BT}+
e^{-(E^{ch}_1(Q_g)+\delta F(T))/k_BT}}\;.
\end{equation}
In Fig.~6 the box charge has been calculated over the main $Q_g$--interval
for a sample of temperatures using Eq.~(\ref{charge2}). There are two
mechanisms how temperature changes the shape of the charge characteristics.
The $k_BT$--terms in Eq.~(\ref{charge2}) make the curve smoother but they
do not shift the critical values of $Q_g$ where the mean charge reach half
integer values. This is just reached by the $\delta F(T)$--term. At
$\Delta/k_BT=8$ corresponding approximately to the cross--over temperature
the 2e--periodicity has already disappeared leaving only the 1e--periodicity.
The property $\langle Q\rangle(Q_g=1/2)=1/2$ means that there are no more
even--odd--effects.

But there is another approach to that problem discussing in more detail
the transition rates~\cite{sch4}. The transition $n=0$ to $n=1$ is
governed by the rate
\begin{equation}
\label{eo}
\Gamma_{1,0}(Q_g)=\frac{1}{e}\mbox{Im}I_{q_s}[-\Delta E^{ch}_{1,0}(Q_g)/\hbar]
\left\{1-e^{\Delta E^{ch}_{1,0}(Q_g)/k_BT}\right\}^{-1}\;.
\end{equation}
This corresponds to $\Gamma_{1,0}$ in Eq.~(\ref{rates0}) for a NIS junction.
The current amplitude $I_{q_s}$ corresponds to the NIS junction and brings
the gap energy $\Delta$ up.
But the back rate describing the transition $n=1$ to $n=0$ consists
of two terms
\begin{equation}
\label{oe}
\Gamma_{0,1}=\gamma_{0,1}+\Gamma^{(a)}_{0,1}\;,
\end{equation}
where the second rate on the right corresponds
again to $\Gamma_{0,1}$ in Eq.~(\ref{rates0}) and reads as
\begin{equation}
\label{oea}
\Gamma^{(a)}_{0,1}(Q_g)=\frac{1}{e}\mbox{Im}I_{q_s}[-\Delta
E^{ch}_{0,1}(Q_g)/\hbar]
\left\{1-e^{\Delta E^{ch}_{0,1}(Q_g)/k_BT}\right\}\;.
\end{equation}
The rate $\gamma_{0,1}$ describes back tunneling of
just the one (odd) electron which does not found any partner for pairing
from the island and is proportional to $N_{eff}^{-1}$~\cite{sch4}. This is
the difference to the approach of Sec.~2. Of course $\gamma_{0,1}$
depends on the energy differences $\Delta E_{0,1}(Q_g)$ too.
The rates are given by ``golden rule '' integrals~\cite{sch4}
\begin{eqnarray}
\label{int1}
\Gamma_{1,0}=\frac{2\pi}{\hbar}\int\limits^{\infty}_{\infty}d\epsilon
N_{(n)}N_{(s)}(\epsilon-\Delta E^{ch}_{1,0})|{\cal T}|^2f(\epsilon)
(1-f(\epsilon-\Delta E^{ch}_{1,0})),\\
\label{int2}
\gamma_{0,1}=\frac{2\pi}{\hbar}\int\limits^{\infty}_{\infty}d\epsilon
N_{(n)}N_{(s)}(\epsilon+\Delta E^{ch}_{0,1})|{\cal T}|^2
(1-f(\epsilon))\tilde{f}(\epsilon+\Delta E^{ch}_{0,1})).
\end{eqnarray}
$N_{(n)}$ and $N_{(s)}$ are the densities of states of normalconductor
and superconductor, respectively. ${\cal T}$ is the tunneling amplitude.
The distribution $\tilde{f}$ is defined by
\[ \int_0^{\infty}d\epsilon N_{(s)}(\epsilon)\tilde{f}(\epsilon)=1\;,\]
corresponding to the assumption that there is just one unpaired electron on
the island.

Now the mean stationary charge reads as ($0<Q_g/e<1$)
\begin{equation}
\label{charge3}
\langle Q\rangle
=e\frac{\Gamma_{1,0}(Q_g)}{\gamma_{0,1}+\Gamma^{(a)}_{0,1}(Q_g)
+\Gamma_{1,0}(Q_g)}\;.
\end{equation}

What is the connection between Eq.~(\ref{charge2}) and Eq.~(\ref{charge3})?
Using Eq.~(\ref{free}), Eq.~(\ref{charge2}) can be transformed in the
following form
\begin{equation}
\label{charge4}
\langle Q\rangle =
e\frac{e^{-(\Delta E^{ch}_{1,0}(Q_G)+\Delta)/k_BT}}{N_{eff}^{-1}+
e^{-(\Delta E^{ch}_{1,0}(Q_g)+\Delta)/k_BT}}\;.
\end{equation}
This expression is a rather good approximation of Eq.~(\ref{charge3})
because for $0<Q_g/e<(1+\lambda)/2$ there is
\[\Gamma_{1,0}(Q_g)\propto\exp(-(\Delta E^{ch}_{1,0}(Q_g)+\Delta)/k_BT))\]
and $\Gamma^{(a)}_{0,1}\ll\gamma_{0,1}$. Also for $(1+\lambda)/2<Q_g/e<1$
this formula yields the correct charge value
($\langle Q\rangle/e\rightarrow 1$). The explicit consideration of the
possibility that just the one unpaired
electron can tunnel back corresponds to the free energy argument in the
other approach.

\section{Superconducting Box with Irradiation}
\label{kap5}

The discussion of the transition rates is necessary because
the treatment of irradiation is not possible in the
Boltzmann distribution approach. But the rates (\ref{eo}, \ref{oe})
have to be generalized with respect to arbitrary $n$ and $Q_g$.
\begin{eqnarray}
\label{ratesn}
\Gamma_{n\pm 1,n}&=&\gamma_{n\pm 1,n}\cdot P_n+\Gamma^{(a)}_{n\pm 1,n}\;,\\
\Gamma^{(a)}_{n\pm 1,n}(Q_g)&=&\frac{1}{e}\mbox{Im}I_{q_s}[-\Delta
E^{ch}_{n\pm 1,n}(Q_g)/\hbar]
\left\{1-e^{\Delta E^{ch}_{n\pm 1,n}(Q_g)/k_BT}\right\}^{-1}\;.
\end{eqnarray}
The rate $\gamma$ in Eq.~(\ref{ratesn}) describes either back tunneling of
just the one (odd) electron (see Sec.~4) or the direct transition of just
the pairing partner of the odd electron on the island.
The other rates $\Gamma$ describe transitions connected with the production
of excited states on the island.

The treatment of the case with microwave irradiation is analogous to that
of Sec.~3. The charge is given by formula~(\ref{charge0}) where the
occupation probabilities $\bar{\sigma}$ are determined by
modified transition rates $\bar{\Gamma}$.
These modified rates have been phenomenological constructed analogous
to Eq.~(\ref{rates1})) and read as
\begin{equation}
\label{rates2}
\bar{\Gamma}_{\ldots}(Q_g)=\sum\limits_rJ_r^2(a)
\Gamma_{\ldots}(\Delta E^{ch}_{\ldots}(Q_g)+r\hbar\bar{\omega})\;,
\end{equation}
where the symbol $\Gamma_{\ldots}$ refers to the rates~(\ref{ratesn}).
Now the occupation probabilities $\bar{\sigma}(n)$ have to be calculated.
This can be done using standard techniques (see e.g.\ Ref.~\cite{gra1}).
Note that $\sum_n\bar{\sigma}(n)=1$. A sample of charge functions with
different AC amplitudes has been plotted in Fig.~7 for fixed temperature.

\section{Discussion}
The investigation shows that the photon assisted tunneling via excited
states leads to an AC amplitude dependent cross--over from the
2e--periodic regime to 1e--periodicity. The main effect is again due
to 1--photon processes ($r=\pm 1$). The critical value of $Q_g$ where the
mean charge reachs the value $1/2$, shows that for fixed temperature
$T<T^{\ast}$ ($T^{\ast}$ is the cross--over temperature in case on no
irradiation) this value goes with increasing microwave amplitude to $1/2$.
This means that the cross--over from 2e--periodic behavior to 1e--periodicity
can be reached not only by increasing temperature but also by increasing
irradiation amplitude. An analytical formula for the cross--over amplitude
cannot be given. The numerical simulation shows that for $T<T^{\ast}$ the
1e--periodic behavior is already established approximately for $a\approx 0.1$.
By increasing the AC amplitude it can also be recognized (cf. Fig.~7) that
the mechanism of ``smoothing'' the staircase function is much more
complicated than that of finite temperature. The graphs for medium AC
amplitudes show kinks at several values $Q_g$ which are relics of the kink
of the NIS current amplitude at the gap energy.

The master equation approach demands that the rates should be much less than
the other relevant frequencies. With respect to the Coulomb frequency
$E_C/\hbar$ this is in principle guaranteed by $R_Q\ll R_T$. But this should
also be ensured with respect to $\bar{\omega}$. Therefore, we have assumed
that $\bar{\omega}$ is approximately of the same order of magnitude as
the Coulomb frequency. If the Coulomb energy corresponds approximately
to 1~K the radiation frequency should be in the range of 20 GHz. It is
reasonable to choose $\bar{\omega}$ in such a way because this yields
the natural physical frequency scale and guarantees more clearly effects.
Up to now there are no experimental measurements known. On the contrary one
tries to suppress photon induced processes by filtering~\cite{vio1}.

\noindent
{\bf Acknowledgments:}
We would like to thank J. Siewert for helpful discussion. For some
useful hints we are indepted to D. Vion.
This work was supported by the Deutsche Forschungsgemeinschaft.

\pagebreak
\begin{figure}
\caption{Scheme of the considered SET-box}
\end{figure}

\begin{figure}
\caption{Charge in the NC box for
different temperatures $E_C/k_BT=100,\;10,\;5,\;1$.
The nearly straight line belongs to  $E_C/k_BT=1$.}
\end{figure}

\begin{figure}
\caption{Charge in the NC box with irradiation at fixed temperature
$E_C/k_BT=100$ for different AC amplitudes
$a=eV_1/\hbar\bar{\omega}=0,\;0.1,\;0.5,\;1$, \quad
($\hbar\bar{\omega}/E_C=1$).}
\end{figure}

\begin{figure}
\caption{The $E_n(V_o)$ parabolas for $\Delta/E_C=0.8$.
The numbers $n$ of the respective parabolas are given by the x--coordinate
of their apices. The lowest crossing points determine the values where
charge jumping occurs for $T=0$ (cf. Fig.~5). Note, that for $\Delta/E_C>1$
the states with odd number $n$ do not participate; only two--electron
tunneling processes (Andreev reflection) take place.}
\end{figure}

\begin{figure}
\caption{Charge in the SC box corresponding
to the case of Fig.~4 for zero temperature.}
\end{figure}

\begin{figure}
\caption{Charge in the SC box in case of $\Delta/E_C=0.8$
at different temperatures $\Delta/k_BT=8,\;10,\;20,\;50,\;100,\;\infty$,
\quad ($N(0){\cal V}\Delta\approx 10^{-4}$ corresponding to the sample of
Ref.~[4]).}
\end{figure}

\begin{figure}
\caption{Charge in the SC box with irradiation in case of
$\Delta/E_C=0.8$ at fixed temperature $\Delta/k_BT=100$ for different
AC amplitudes $a=eV_1/\hbar\bar{\omega}=0,\;0.01,\;0.02,\;0.05,\;0.1$,
\quad ($\hbar\bar{\omega}/E_C=1$ , $N(0){\cal V}\Delta\approx 10^{-4}$
corresponding to the sample of Ref.~[4]).}
\end{figure}

\end{document}